\documentclass[conference]{IEEEtran}
\IEEEoverridecommandlockouts
\usepackage{cite}
\usepackage{amsmath,amssymb,amsfonts}
\usepackage{graphicx}
\usepackage{textcomp}
\usepackage{xcolor}
\usepackage{multirow}
\usepackage{makecell}
\usepackage[draft=true]{hyperref}
\usepackage[ruled,vlined]{algorithm2e}
\usepackage[utf8]{inputenc}
\usepackage[english]{babel}
\usepackage{lastpage}
\usepackage{adjustbox}
\usepackage{fancyhdr}
\usepackage{balance}
\usepackage{booktabs} 
\usepackage[left=2cm, right=2cm, top=1.8cm, bottom=2.7cm]{geometry}
\makeatletter
\renewcommand{\fnum@figure}{Fig. \thefigure}
\makeatother

\begin{document}

\title{
HierSFL: Local Differential Privacy-aided Split Federated Learning in Mobile Edge Computing
\\
}

\author{
        \IEEEauthorblockN{
        Minh K. Quan\IEEEauthorrefmark{1},
		Dinh C. Nguyen\IEEEauthorrefmark{2}, Van-Dinh Nguyen\IEEEauthorrefmark{3}, 
	}
        \IEEEauthorblockN{
        Mayuri Wijayasundara\IEEEauthorrefmark{1}, Sujeeva Setunge\IEEEauthorrefmark{4},
        Pubudu N. Pathirana\IEEEauthorrefmark{1}
	}
	\IEEEauthorblockA{\IEEEauthorrefmark{1}School of Engineering, Deakin University, Australia \\
	\IEEEauthorrefmark{2}Department of Electrical and Computer Engineering, University of Alabama in Huntsville, USA  \\
	 \IEEEauthorrefmark{3}College of Engineering and Computer Science, VinUniversity, Vinhomes Ocean Park, Hanoi 100000, Vietnam \\
	 \IEEEauthorrefmark{4}School of Engineering, Royal Melbourne Institute of Technology University, Melbourne, VIC 3000, Australia
	}}
	\markboth{}%
	{}

\maketitle
\pagenumbering{gobble} 
\begin{abstract}
     Federated Learning is a promising approach for learning from user data while preserving data privacy. However, the high requirements of the model training process make it difficult for clients with limited memory or bandwidth to participate. To tackle this problem, Split Federated Learning is utilized, where clients upload their intermediate model training outcomes to a cloud server for collaborative server-client model training. This methodology facilitates resource-constrained clients' participation in model training but also increases the training time and communication overhead. To overcome these limitations, we propose a novel algorithm, called Hierarchical Split Federated Learning (HierSFL), that amalgamates models at the edge and cloud phases, presenting qualitative directives for determining the best aggregation timeframes to reduce computation and communication expenses. By implementing local differential privacy at the client and edge server levels, we enhance privacy during local model parameter updates. Our experiments using CIFAR-10 and MNIST datasets show that HierSFL outperforms standard FL approaches with better training accuracy, training time, and communication-computing trade-offs. HierSFL offers a promising solution to mobile edge computing's challenges, ultimately leading to faster content delivery and improved mobile service quality.

\end{abstract}


\section{Introduction}
\label{Section:Introduction}
The burgeoning prevalence of interconnected devices has given rise to a considerable upswing in the generation of disparate data, encompassing not only smart phones but also smart watches and fitness monitors \cite{wang2019big}. The intricacies entailed in acquiring knowledge related to data privacy preservation, in addition to the prohibitive costs of transmitting all data to a remote cloud \cite{nath2020deep}, have facilitated the emergence of a potential solution, mobile edge computing (MEC). The adoption of MEC technology offers several prospective benefits, including heightened energy efficiency and decreased latency. Notwithstanding, this approach also raises several privacy-related apprehensions, such as inadvertent disclosure of sensitive user information, interception of trained models, and an inadequate regulatory frameworks \cite{wang2018edge}. Consequently, a robust and holistic approach to privacy preservation and governance is imperative for the sustainable proliferation of MEC technology.

To address these MEC-related privacy issues \cite{wang2019edge}, federated learning (FL) is commonly regarded as an effective framework. This is attributed to its capacity to facilitate learning procedures using local data, while also preserving sensitive information. The conventional FL framework enables users to train their learning models on local data after which they send only the necessary trained weights to a central server for aggregation. Once the local models are aggregated, the updated global model is broadcast back to the users for further refinement. While FL provides protection for users' privacy, it demands significant resources for both communication and model training, which could be quite challenging for low-cost devices. The process of transmitting model updates iteratively to the central cloud server creates a substantial communication load that can be quite challenging to manage \cite{li2021hermes}. Moreover, training of complex deep neural networks (DNN) imposes a significant demand on memory and computational resources \cite{deng2021share}. These obstacles can impede the widespread incorporation of FL, particularly for limited-resource mobile clients.

To cope with clients' resource limitation issues in FL, the concept of split FL (SFL) has been proposed. \cite{ thapa2022splitfed}. The key benefit of SFL is that it allows elaborating models to be learned on mobile devices without overburdening them. The SFL methodology involves partitioning of the machine learning (ML) model into several segments, with one of these segments being designated to support remote server training. The clients participate in the propagation of their models, while concurrently transmitting intermediate outcomes to a remote server to facilitate the completion of model training. The incorporation of parallel processing and network splitting in SFL's training strategy results in reduced memory and processing requirements for clients when compared to FL, though certain limitations must still be addressed. Unlike the challenges faced in FL, the frequent transfer of partial results to the remote server poses a significant impediment, as it amplifies communication resource usage and transmission time, and calls for measures to guarantee the confidentiality of intermediate results during transmission.

\subsection{Motivation and Key Contributions}
Since its introduction in \cite{thapa2022splitfed}, SFL has undergone continuous enhancements and has been integrated into various research endeavours. The study \cite{turina2020combining} examined two learning architectures that merge FL and  split learning (SL) to lessen client computational demands and parallelize SL. In contrast, \cite{liu2022wireless} implemented SFL in  unmanned aerial vehicle (UAV) networks to address data transfer and privacy concerns. The authors \cite{yang2022robust} also highlighted an SFL application in U-shaped medical image networks. However, SFL faces several challenges, including central cloud server strain, communication delays, and privacy concerns. SFL relies on a central server to aggregate updates from all clients. This can lead to bottlenecks and performance degradation, especially with many clients. Additionally, all clients must communicate with the central server at each iteration, which can cause communication delays, especially for remote clients or congested networks. Finally, SFL requires clients to send their local model updates to the central server frequently without privacy-protecting techniques, which raises privacy concerns, as attacks can occur during communication between the server and clients.

To address these challenges, we propose $\mathsf{HierSFL}$, a novel framework that utilizes mobile edge servers (MESs) as training assistants and model aggregators. In $\mathsf{HierSFL}$, clients are divided into groups and assigned to MESs. Clients in each group do not need to send their data to the central server; instead, they use local differential privacy (LDP) to add noise to their data before sending it to their assigned MES. MESs then aggregate the updates from their assigned clients and send the aggregated updates to the central server. The central server updates the global model and sends it back to the MESs, which distribute it to their assigned clients. This approach significantly reduces the load on the central server and improves communication delays, as clients only need to communicate with their assigned MES, which is typically located closer to them. Our contributions can be summed up as follows:

\begin{itemize}
    \item We propose a hierarchical SFL framework with model aggregation at both the MES and cloud levels, where qualitative guidelines are developed to determine optimal aggregation intervals at each level. This helps balance computation and communication costs.
    \item  We implement LDP at both the client-level model to enhance confidentiality during the synchronization of local model parameters.
    \item We conduct experiments using the CIFAR-10 and MNIST datasets, which demonstrate the superiority of $\mathsf{HierSFL}$ scheme over conventional FL approaches with better communication-computing trade-offs.
\end{itemize}

\subsection{Paper Organization}
The following sections of this study are organized as such. In details, Section \ref{Section:SystemDesign} explicates the learning problem and elucidates the $\mathsf{HierSFL}$ framework. Section \ref{Section:SimulationAndEvaluation} is dedicated to presenting the results of the simulations conducted for $\mathsf{HierSFL}$, which serve the dual purpose of validating the convergence analysis and showcasing the benefits of the $\mathsf{HierSFL}$ approach. In the final section of the paper, Section \ref{Section:Conclusion}, a summary of the main findings and contributions is presented.

\section{System Design}
\label{Section:SystemDesign}

In this section, the primary learning problems in FL and SFL are discussed, and $\mathsf{HierSFL}$, a three-tier FL system, is presented as a solution to enable SFL in a hierarchical MEC network followed by an introduction of the LDP mechanism for enhancing privacy in $\mathsf{HierSFL}$.

\subsection{Learning Problem Overview}
\subsubsection{Federated Learning (FL)}
In the context of FL learning approach for training ML models, each client has its own dataset represented by $\mathcal{S}_k={(x_{k,i}, y_{k,i})}_{i=1}^{|\mathcal{S}_k|}$, where $x_{k,i}$ represents the $i$-th input feature vector and $y_{k,i}$ represents its corresponding target value. The model is characterized by a real vector $\boldsymbol{w}$, while the overall dataset $\mathcal{S}$ is constructed by merging the individual datasets $\mathcal{S}_k$ for $k=1,2,\dots,K$, resulting in a total of $|\mathcal{S}|$ samples. The prediction error for the $i$-th sample is given by the loss function $\ell(x_{k,i}, y_{k,i}, \boldsymbol{w})$ or $\ell_{k,i}(\boldsymbol{w})$. The objective of FL is to minimize the empirical risk function:
\begin{equation}
    \label{eq:FL_global_loss}
    L(\boldsymbol{w}) = \frac{\sum_{k=1}^{K} \sum_{i \in \mathcal{S}_k} \ell_{k,i}(\boldsymbol{w})}{|\mathcal{S}|},
\end{equation}
which is based on non-IID (Non-Independent and Identically Distributed) datasets of different clients \cite{sun2019survey}. The function $L(\boldsymbol{w})$ specified in equation \eqref{eq:FL_global_loss} serves as the global loss function. 

Theoretically, the convexity of $L(\boldsymbol{w})$ depends on whether the individual loss functions $\ell_i(\boldsymbol{w})$ and their aggregation are convex. However, in practice, FL optimization problems are often non-convex, posing a challenge for Gradient descent and its variants that rely on convexity to find the global optimum. To address this, stochastic gradient descent (SGD) and its variations have been employed for optimizing $L(\boldsymbol{w})$ by randomly selecting a mini-batch $\mathcal{B}_{t}$ of samples $\mathcal{B}$ from the training data at each iteration $t$. The gradient descent algorithm involves updating the parameter vector $\boldsymbol{w}$ by moving in the direction opposite to the gradient. To efficiently compute the loss function's gradient with the parameter vector $\boldsymbol{w}$, it is possible to adopt the following equation:
\begin{equation}
    \boldsymbol{w}_{t+1} = \boldsymbol{w}_{t} - \eta_{t} \frac{\sum_{(x,y)\in \mathcal{B}_{t}} \nabla \ell(x,y,\boldsymbol{w})}{|\mathcal{B}_{t}|}.
\end{equation}
The SGD algorithm samples a data point $(x,y)$ randomly from the dataset $S$ and updates the parameter vector $\boldsymbol{w}$ using the learning rate $\eta_t$ and gradient $\nabla \ell(x,y,\boldsymbol{w})$ of the loss function at iteration $t$ pertaining to $\boldsymbol{w}$. The algorithm continues to iterate until reaching a desired level of convergence or a maximum number of iterations.

\subsubsection{Split Federated Learning (SFL)}
In the SFL framework, $ \mathcal{S}t $ comprises a group of $K$ clients at time $t$. Each client performs parallel forward propagation for their models with a noise layer. They then send compressed data $D_{k,t}$ and labels $Y_k$ to the central server. $p_k$ is the sample size for client $k$, $p$ is the total sample size. During training, client $k$ interacts with both $\rho_1$ and $\rho_2$ servers. Specifically, at server $\rho_1$, the subsequent steps are executed:
\begin{enumerate}
    \item  The forward propagation procedure on the global on the server model $\boldsymbol{w}^s_t$ with $D_{k,t}$.
    \item Computation of predicted labels $\hat{Y_k}$.
    \item Loss calculation with $Y_k$ and $\hat{Y_k}$ based on the following equation: 
    \begin{equation}
        \label{eq:lossFunctionSFL}L(\boldsymbol{w}^s_t;D_{s,t})=\sum_{k=1}^K\frac{p_k}{p}\ell(Y_k, \hat{Y_k}),
    \end{equation}
    where $n$ is the total sample size, and $n_k$ is the sample size of client $k$.
    \item Separate parallel processing of compressed data from each client during back-propagation on the server-side model. Respective clients receive the gradients of compressed data $\nabla \ell_k(\boldsymbol{w}^s_t ; D_{s,t})$ for use in their back-propagation process.
    \item The server's model is updated via FedAvg, which involves taking a weighted average of the gradients computed during back-propagation on each client's compressed data: 
    \begin{equation}
        \boldsymbol{w}^s_{t+1} = \boldsymbol{w}^s_t - \eta_t \frac{1}{p} \sum_{k=1}^K \frac{p_k}{p} \nabla \ell_k(\boldsymbol{w}^s_t; D_{s,t}).
    \end{equation}
\end{enumerate}

After receiving the gradients of its compressed data $\nabla \ell_k(\boldsymbol{w}^s_t ; D_{s,t})$, each client utilizes them to conduct back-propagation on their local model and derive its gradients $\nabla \ell_k(\boldsymbol{w}^c_{k,t})$. The gradients are secured via a LDP mechanism before being sent to the server $\rho_2$, which performs a FedAvg of client-side local updates and disseminates the results to all participating clients for privacy:
\begin{equation}
    \boldsymbol{w}^c_{t+1} = \frac{1}{p} \sum_{k=1}^K \frac{p_k}{p} \boldsymbol{w}^c_{k,t}.
\end{equation}

\subsection{HierSFL Framework}
\begin{figure}[htbp]
\centerline{\includegraphics[height=1.25\linewidth, width=0.96\linewidth]{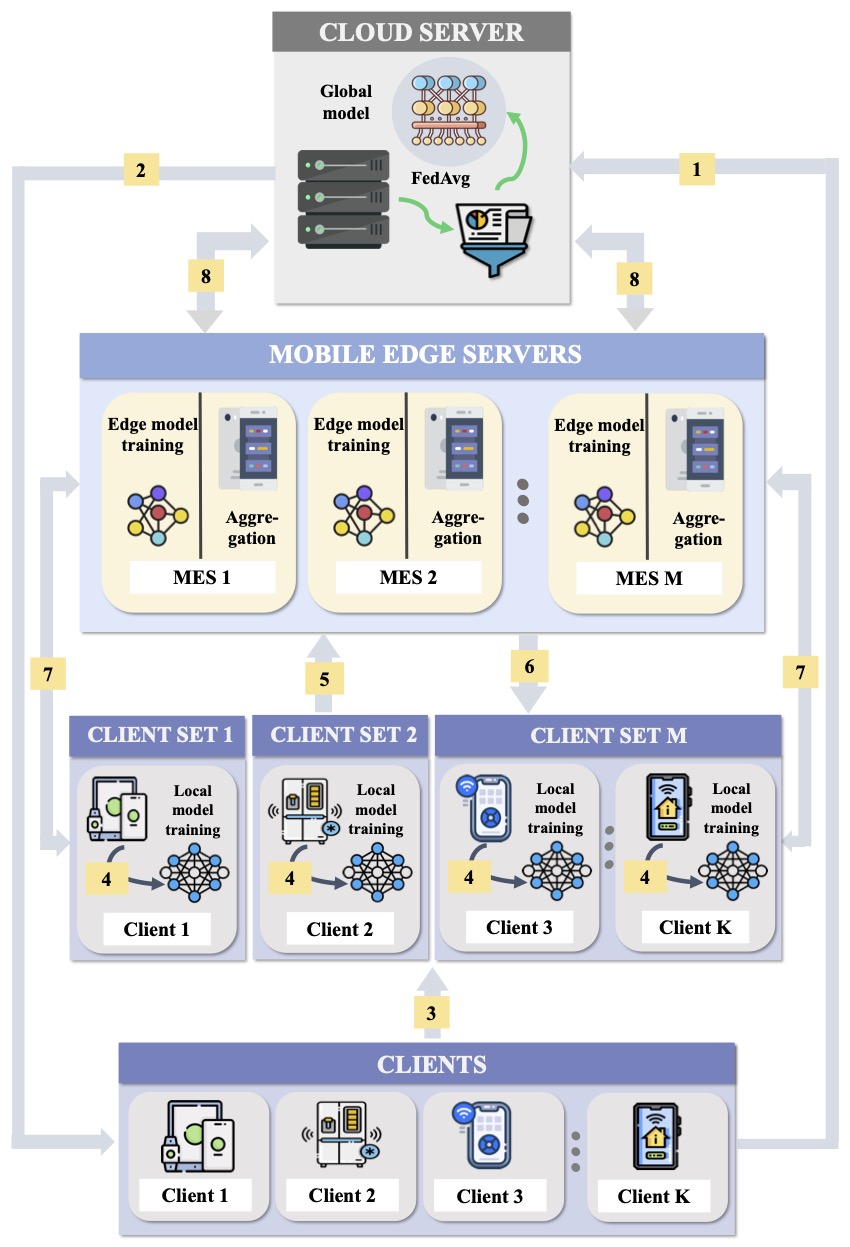}}
\caption{The workflow of $\mathsf{HierSFL}$ framework.}
\label{fig1}
\end{figure}
As discussed in the previous section, SFL is a novel approach to ML that reaps the benefits of FL and SL. While model aggregation on the cloud parameter server can accommodate a large number of customers, it incurs substantial communication expenses. In contrast, a small number of clients participating in model aggregation at the MEC parameter server leads to a drastic reduction in communication expenses. Consequently, a $\mathsf{HierSFL}$ framework is proposed to reap advantages of both approaches. The framework that we propose comprises a cloud server and $\mathcal{M}$ MESs, each identified by the index $m$. MESs serve separate client sets, which are labeled as ${\{C^m\}}^\mathcal{M}_{m=1}$. Furthermore, there are $K$ clients, indexed by both $k$ and $m$, each possess distributed datasets ${\{D^m_k\}}^K_{k=1}$. The dataset collected under each MES is represented as $D^m$. It is the responsibility of each MES to facilitate collaborative training and aggregation of models from the clients that it serves.

The working flow of our proposed $\mathsf{HierSFL}$ scheme is presented in Fig. \ref{fig1}. Specifically, from Step 1 to 3, the central cloud server uses client information to assign clients to MESs. Before the collaborative training steps including 5 and 6, clients undertake local model training in Step 4. Afterwards, in Steps 5 and 6, clients work together on parallel model training, keeping the MES updated with intermediate results and receiving edge-side assistance and gradients for further advancement. This cycle occurs for $E$ iterations. In Step 7, MES $m$ aggregates model parameters using the FedAvg algorithm after every $p_1$ local update for each client in the $C_m$ set. The edge aggregation procedure concludes after $p_2$ rounds, and model parameters are forwarded from the MESs to the cloud server (Step 8). The cloud server employs the FedAvg algorithm to combine the model parameters and dispatch them to MESs, which update the edge model and then broadcast it to clients. The $\mathsf{HierSFL}$ algorithm supports collaborative machine learning while preserving data privacy and minimizing communication overhead. A thorough description of this workflow is available in Algorithm \ref{alg:HierSFL}.

\subsection{Local Differential Privacy (LDP) in $\mathsf{HierSFL}$}
To preserve the privacy of client data in $\mathsf{HierSFL}$, where the deep neural network is trained collaboratively by the cloud, MESs, and clients, LDP \cite{arachchige2019local} can be implemented. From a mathematical standpoint, let $x$ denote the output of a layer in the client local model, and $\sigma(x)$ represent the function that adds calibrated noise to $x$. The output $y = \sigma(x)$ is said to satisfy $\varepsilon$-LDP if the following inequality holds for all neighboring input pairs $x$ and $x'$, as well as for all sets of outputs $M$:
\begin{equation}
    \mathrm{Pr}[\sigma(x) \in S] \leq e^{\varepsilon} \cdot \mathrm{Pr}[\sigma(x') \in S].
\end{equation}

To preserve privacy in the context of LDP, the local client model weights can be perturbed by introducing calibrated noise. Let $\boldsymbol{w}$ represent the original weight vector, and $\boldsymbol{w}'$ denote the perturbed weight vector that satisfies $\varepsilon$-LDP. The perturbed weight vector can be obtained by adding noise drawn from a Laplace distribution \cite{wu2020privacy} to the original weights:
\begin{equation}
\label{eq:laplace}
\boldsymbol{w}' = \boldsymbol{w} + \boldsymbol{\delta},
\end{equation}
where $\boldsymbol{\delta}$ is a noise vector drawn from a Laplace distribution with scale parameter $c$. The scale parameter $c$ is determined based on the sensitivity of the weights and the desired privacy parameter $\varepsilon$: 
\begin{equation}
    c = \frac{\Theta_{\boldsymbol{w}}}{\varepsilon}.
\end{equation}
The sensitivity of the weights, denoted as $\Theta_{\boldsymbol{w}}$, captures the maximum possible change in the weight vector when any single individual's data is modified. It can be defined as:
\begin{equation}
\label{eq:sensitivity}
\Theta_{\boldsymbol{w}} = \max_{\boldsymbol{w}, \boldsymbol{w}'} \|\boldsymbol{w} - \boldsymbol{w}'\|_1,
\end{equation}
where $\|\cdot \|_1$ represents the $L_1$ norm. Incorporating Laplace noise into the $L_1$ norm adds random changes, enhancing individual contribution privacy. The noise level is determined by the privacy budget $\varepsilon$, with lower $\varepsilon$ values offering stronger privacy protection but potentially reducing accuracy.


\begin{algorithm}
\caption{Proposed $\mathsf{HierSFL}$ Framework}
\label{alg:HierSFL}
\SetKwBlock{void}{void}{}
\SetKwBlock{function}{function}{}
\SetKwInOut{Input}{Input}
\SetKwInOut{Output}{Output}

\textbf{Input:}
Total aggregation rounds: $P$. 
Local updates and edge model aggregations per client: $p_1$, $p_2$. 
Local training epoch: $E$.
Clients served by MES $m, \forall m \in \mathcal{M}$: $C_m$.

\textbf{Initilaization:}
Initialize the global client model $\boldsymbol{w}^\ell_0$, the MES model $\boldsymbol{w}^m_0$. Send the models to clients and MESs. 
\BlankLine

\For{aggregation round $p\leftarrow 1$ \KwTo $P$} {
\BlankLine
\For{each MES $m \in \mathcal{M}$, client $k \in C_m$ parallel}{ 
    \For{epoch $e\leftarrow 1$ \KwTo $E$}{
        \textit{/* Client forward propagation */}\\
        Calculate $\theta^k_p, Y_k$ in $\boldsymbol{w}^\ell_k(p)$;
        \BlankLine
        \textit{/* MES forward propagation and back propagation */}\\
        Calculate $\nabla f_k(\boldsymbol{w}^m(p))$ based on $Y_k$ and $\hat{Y_k}$;
        \BlankLine
    
        \textit{/* Client back propagation */}\\
        Update $\boldsymbol{w}^\ell_k(p)$ using $\nabla f_k(\boldsymbol{w}^m(p))$;\\
        Perturb weights $\boldsymbol{w}^\ell_k(p)$ as: $\boldsymbol{w}^\ell_k(p) \leftarrow \boldsymbol{w}^\ell_k(p) + \boldsymbol{\delta}$;
        }
    }
}
\BlankLine

\If{$p$ mod $p_1$ = 0}{
\For{each MES $m$ parallel}{
\textit{/* MES aggregation */}\\
$\boldsymbol{w}^m \leftarrow \frac{1}{|D^m|} \sum_{k \in C_m} |D^m_k| {\boldsymbol{w}^\ell_k(p)}_{k \in C_m}$;\\
\If{$p$ mod $p_1.p_2 \neq 0$}{
Update MES clients: $\boldsymbol{w}^m_k(p) \leftarrow \boldsymbol{w}^m(p)$;\\}}}

\BlankLine
\If{$p$ mod $p_1.p_2 = 0$}{
\textit{/* Cloud aggregation */}\\
$\boldsymbol{w}(p)$ $\leftarrow$ $\frac{1}{|D|} \sum_{m \in \mathcal{M}} |D^m| {\boldsymbol{w}^m(p)}_{m \in \mathcal{M}}$;\\
Update all clients: $\boldsymbol{w}^m_k(p) \leftarrow \boldsymbol{w}(p)$;\\}

\textbf{Output:}
The final global model $\boldsymbol{w}(p)$ after $P$ aggregation rounds
\end{algorithm}

\section{Simulations and Evaluation}
\label{Section:SimulationAndEvaluation}

\subsection{Simulation Settings}
We conducted experiments on a server equipped with a 10-core CPU, 8-core GPU, and 16 GB of RAM. Our system was built using PyTorch 1.10.0, and we simulated server-client transmission delays using the Python \texttt{time.sleep()} function. To evaluate the system, we considered different combinations of client sets ($k \in {20, 40, 60, 80}$) and MES sets ($m \in {4, 8, 12, 16}$). Each MES authorized an equal number of clients with the same volume of training data. In scenarios involving data distribution, each client was assigned two distinct sample labels, each containing 400 samples, ensuring a non-IID data distribution.

In our assessment of the $\mathsf{HierSFL}$ framework, we conduct image classification tasks employing the commonly used CIFAR-10 and MNIST datasets, ensuring an optimal selection for the evaluation. For CIFAR-10, we employ a CNN model with 5,852,170 parameters, including an output layer and $\mathsf{3 \times 3}$ convolution layers, trained using 50,000 samples of $\mathsf{32 \times 32}$ pixel images. Similarly, for MNIST, we use a CNN model with 21,840 parameters \cite{mcmahan2017communication}, trained with 60,000 samples of 10-class handwritten images measuring $\mathsf{28 \times 28}$ pixels. Both datasets undergo local computation with an initial learning rate of $\mathsf{\eta = 0.01}$, an exponential learning rate decay of $\mathsf{\varphi = 0.995}$ per epoch, Stochastic Gradient Descent momentum $\mathsf{\tau=0.5}$, two different privacy budgets $\varepsilon_1 = 0.5$ (MNIST) and $\varepsilon_2 = 5$ (CIFAR-10), and a batch size of $b = 32$, providing a robust evaluation of the $\mathsf{HierSFL}$ framework. Furthermore, we consider the three existing baselines, including conventional FL \cite{li2019convergence}, SFL \cite{thapa2022splitfed}, and Hierarchical FL (HFL)  \cite{liu2020client},
respectively.

\subsection{Simulation Results}
\begin{figure}[htbp]
\centerline{\includegraphics[width=0.99\linewidth]{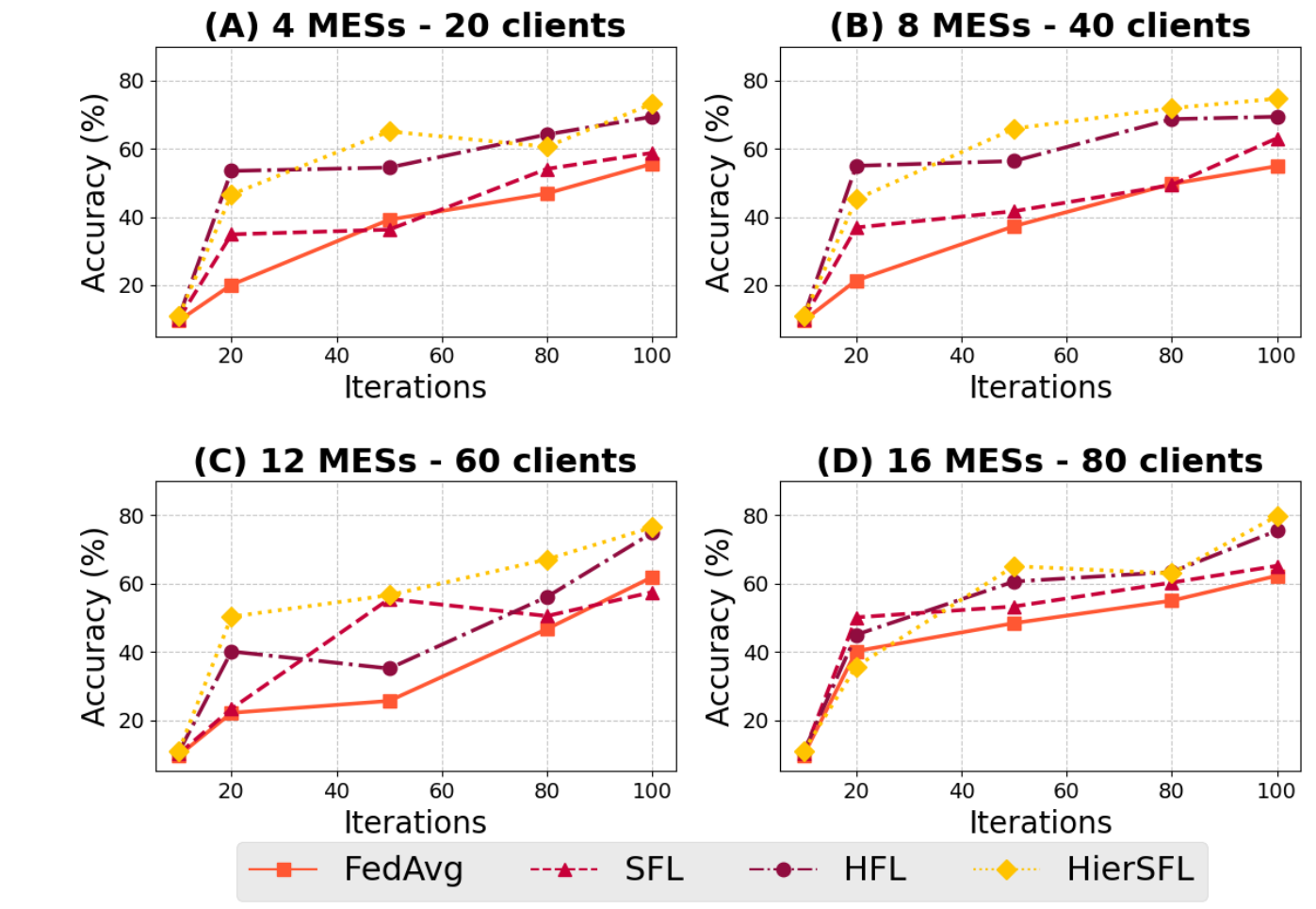}}
\caption{Fine-tuning MES and client numbers for optimal CIFAR-10 accuracy.}
\label{fig3}
\end{figure}

We compared $\mathsf{HierSFL}$ to FL, SFL, and HFL using ResNet18 on CIFAR-10, varying MESs from 4 to 16 and clients from 20 to 80. We assessed training accuracy after a fixed number of iterations, with $p_1$ set to 5 and $p_2$ to 2. In most cases, $\mathsf{HierSFL}$ outperformed, achieving 79.8\% accuracy with 16 MESs and 80 clients, surpassing FL (62.3\%), SFL (65.2\%), and HFL (75.5\%) as shown in Fig. \ref{fig3}. However, as network complexity increased, computational demands on FL and HFL clients grew. While SFL's server-client approach helped, it could strain the cloud server with a larger number of clients. Thus, $\mathsf{HierSFL}$'s hierarchical design optimizes network topology and improves learning in such cases. However, in the 4-edge-server and 20-client setting, HFL achieved higher accuracy (64.2\%) than $\mathsf{HierSFL}$ (60.6\%) after 80 iterations, underscoring the role of network topology, dataset properties, and algorithm design in learning outcomes. Additionally, the performance of FL and SFL varied in different network configurations. For instance, with 8-edge-servers and 40 clients, SFL achieved 63.1\% accuracy after 100 iterations with ResNet18, while FL only reached 54.9\%. With 12 MESs and 60 clients, FL outperformed SFL with an accuracy of 61.9\% compared to SFL's 57.5\%.

\begin{figure}[htbp]
\centerline{\includegraphics[width=0.96\linewidth]{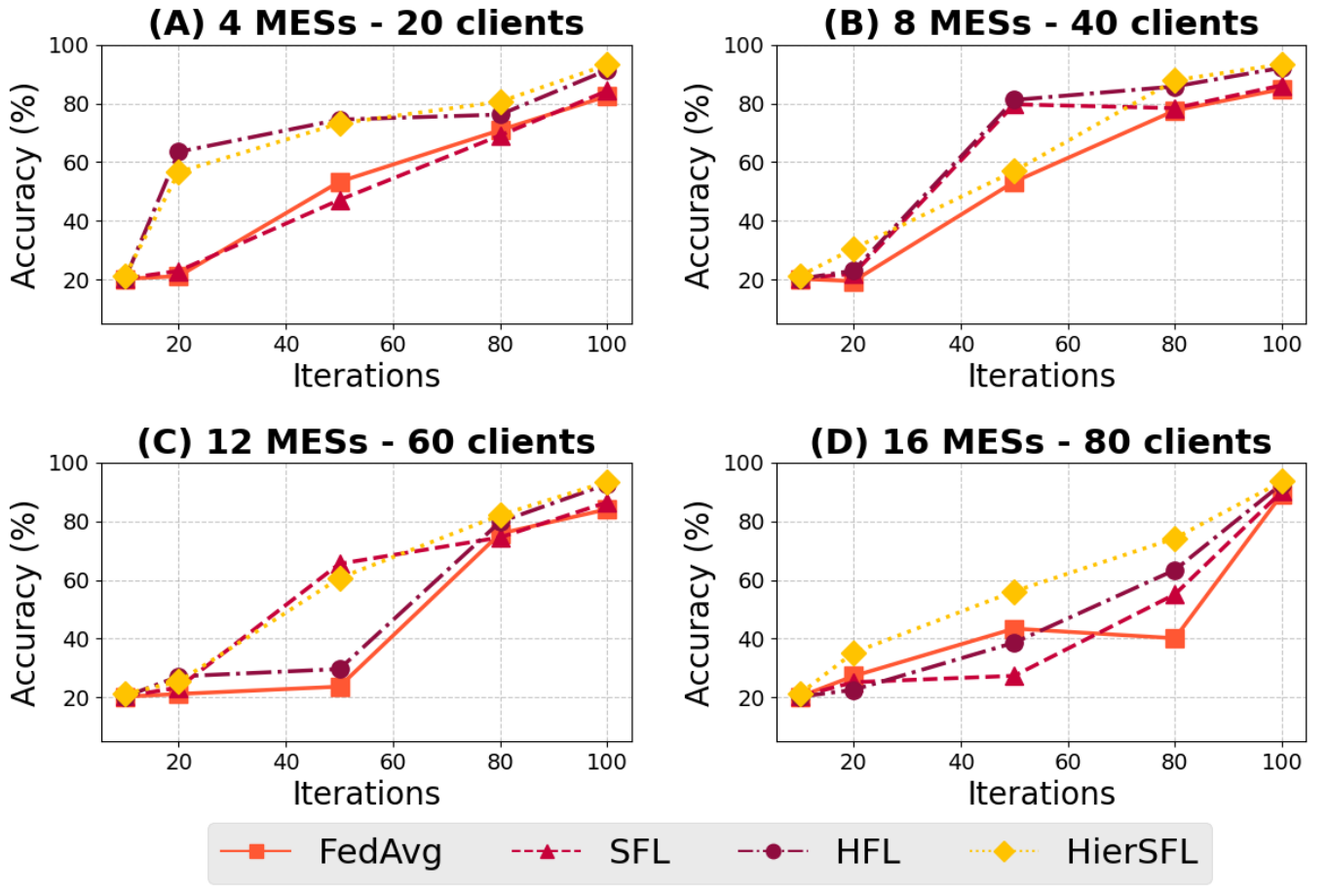}}
\caption{Fine-tuning MES and Client Numbers for Optimal MNIST Accuracy.}
\label{fig4}
\end{figure}

\begin{figure}[htbp]
\centerline{\includegraphics[width=0.95\linewidth]{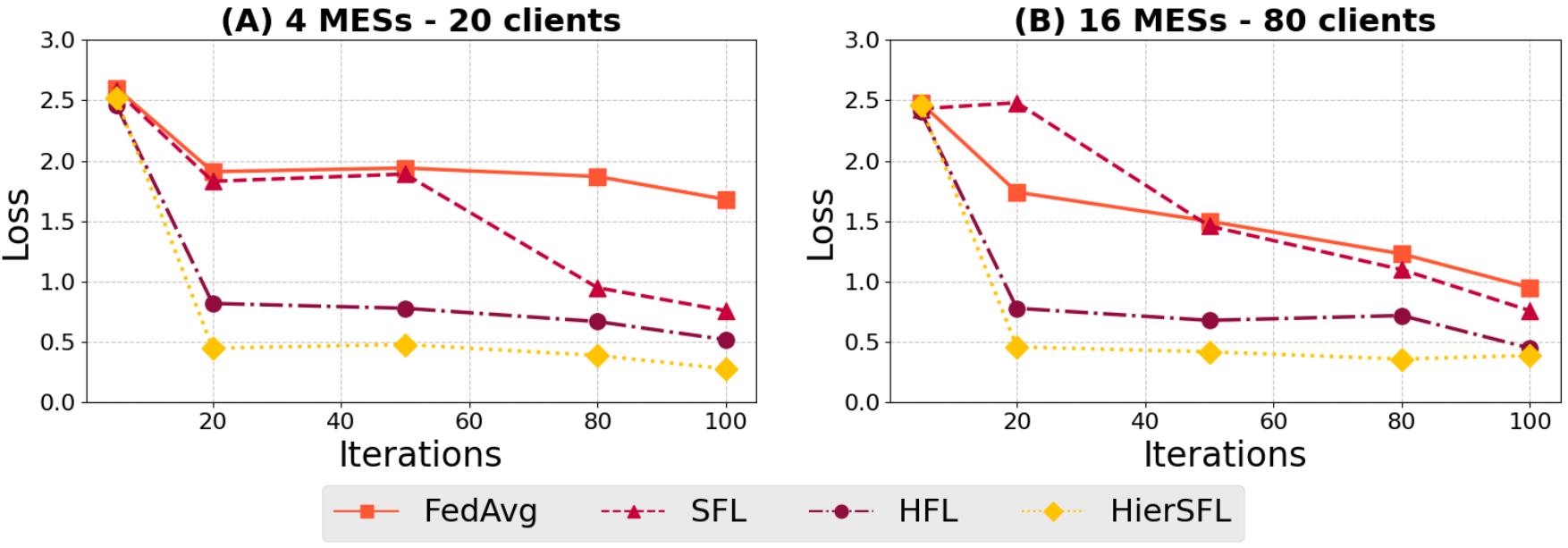}}
\caption{Loss values comparison of FL, SFL, HFL and $\mathsf{HierSFL}$ on MNIST dataset.}
\label{fig5}
\end{figure}

Using the same client and MES count as in the CIFAR-10 evaluation, we tested $\mathsf{HierSFL}$ with MNIST and ResNet50 in four scenarios. As shown in Fig. \ref{fig4}, $\mathsf{HierSFL}$ outperformed FL, SFL, and HFL in accuracy across all scenarios, with a maximum gain of 1.8\% and a minimum of 0.8\% after 100 iterations. It also showed faster convergence in most cases. In terms of accuracy, $\mathsf{HierSFL}$ achieved 93.2\%, 93.4\%, 93.5\%, and 93.6\%, while HFL came second with accuracies between 91.4\% and 92.8\%. FL and SFL recorded accuracies below 90\% in all scenarios. This highlights that $\mathsf{HierSFL}$ and HFL are the most effective methods for faster convergence. Fig. \ref{fig5} compares convergence rates. In the 4-edge, 20-client setup (Subplot A), $\mathsf{HierSFL}$ achieved the fastest convergence with a 0.28 loss after 100 iterations. HFL followed closely with a 0.52 loss, while SFL and FL had higher losses. In the 16-edge, 80-client configuration (Subplot B), $\mathsf{HierSFL}$ maintained its lead, with HFL close behind, further illustrating their superior convergence rates.

\begin{figure}[htbp]
\centerline{\includegraphics[width=0.98\linewidth]{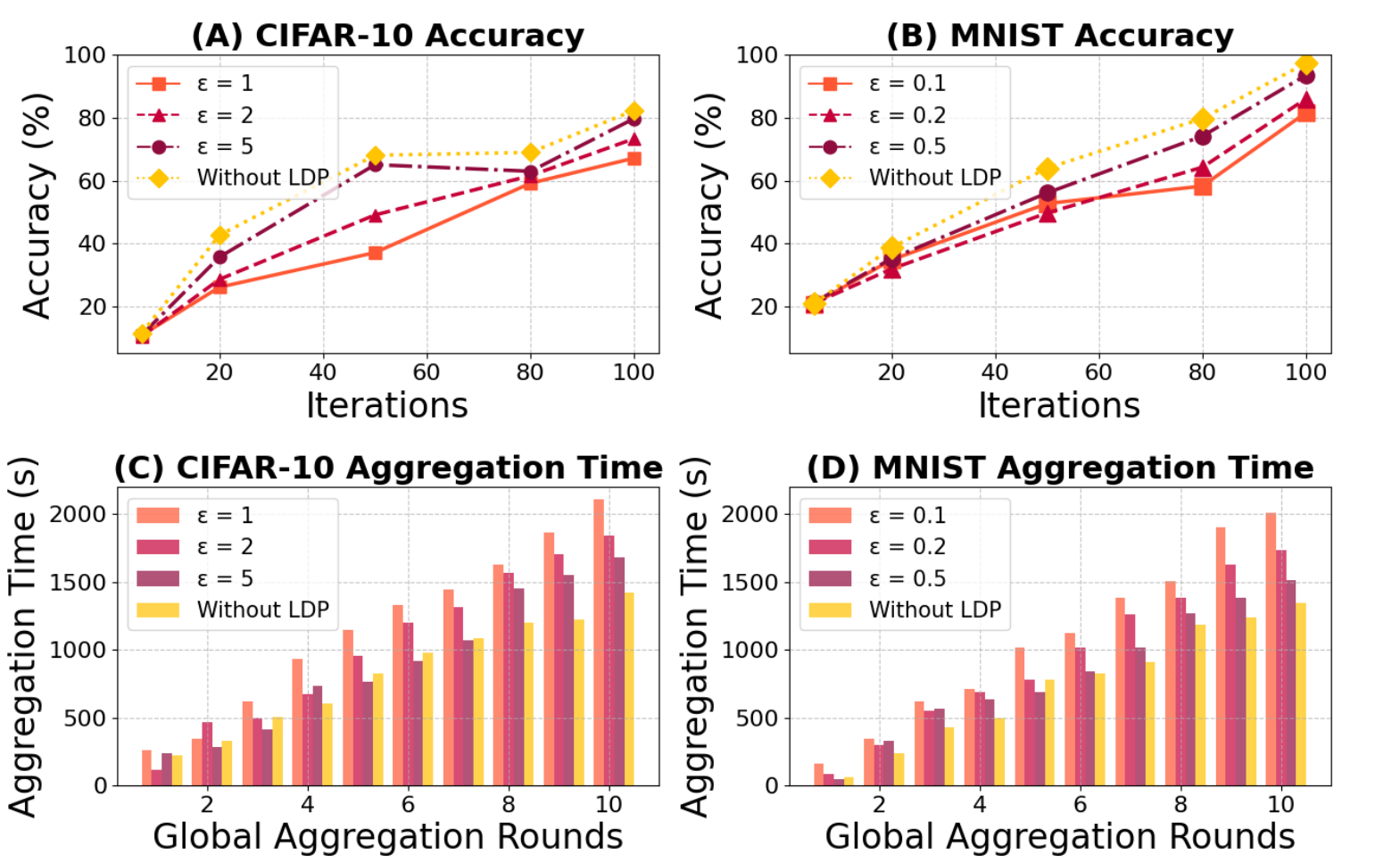}}
\caption{Impact of privacy budget $\varepsilon$ on training accuracy and aggregation time for $\mathsf{HierSFL}$ with 4 MESs and 20 clients.}
\label{fig6}
\end{figure}

In Fig. \ref{fig6}, we illustrated a prominent trade-off between privacy, as quantified by the privacy budget $\varepsilon$, and the training accuracy of our HierSFL method on both CIFAR-10 and MNIST datasets. The choice of $\varepsilon$ values was deliberate, with larger values (1, 2, and 5) selected for CIFAR-10 and smaller values (0.1, 0.2, and 0.5) for MNIST. This distinction is driven by the datasets' characteristics; CIFAR-10, featuring more intricate and sensitive image data, benefits from larger $\varepsilon$ values to enhance model accuracy. In contrast, the simpler MNIST dataset requires stronger privacy protection, hence smaller $\varepsilon$ values. These $\varepsilon$ choices facilitate meaningful comparisons within the field, underscoring the adaptability of our method to diverse privacy needs. Notably, training without LDP, where $\varepsilon$ is not applied, achieves the highest accuracy but comes at the expense of privacy. Inverting $\varepsilon$ values could lead to increased accuracy for CIFAR-10 at the expense of privacy, while MNIST may favor privacy over accuracy due to added noise. This understanding enhances meaningful comparisons between HierSFL and other methods across different privacy scenarios.

\textit{Performance Comparison with SFL and Hierarchical FL:} Comparing the training times of $\mathsf{HierSFL}$, HFL, and SFL using CIFAR-10 (Table \ref{tab:cifar_training_time}) and MNIST (Table \ref{tab:mnist_training_time}) datasets, the tables list durations in seconds for 10, 15, 20, and 25 global aggregation rounds.
\begin{table}[ht]
\renewcommand{\arraystretch}{1.2}
\centering
\caption{CIFAR-10 Dataset Training Duration (seconds).}
\begin{tabular}{|c|c|c|c|c|}
\hline
\multirow{2}{*}{\textbf{Framework}} & \multicolumn{4}{c|}{\textbf{Global Aggregation Rounds}} \\ \cline{2-5}
  &  10     & 15     &  20      & 25      \\ 
\hline
SFL             & 1735.06 & 2635.66 & 3258.83 & 3868.33 \\ 
\hline
HFL              & 1724.14 & 2556.68 & 3105.96 & 3690.21 \\
\hline
\textbf{$\mathsf{HierSFL}$}     & \textbf{1680.51} & \textbf{2420.24} & \textbf{2924.11} & \textbf{3502.75} \\
\hline
\end{tabular}
\label{tab:cifar_training_time}
\end{table}

\begin{table}[ht]
\renewcommand{\arraystretch}{1.2}
\centering
\caption{MNIST Dataset Training Duration (seconds).}
\begin{tabular}{|c|c|c|c|c|}
\hline
\multirow{2}{*}{\textbf{Framework}} & \multicolumn{4}{c|}{\textbf{Global Aggregation Rounds}} \\ \cline{2-5}
  &  10     & 15     &  20      & 25      \\ 
\hline
SFL             & 1635.21 & 2302.16 & 3012.83 & 3629.15 \\ 
\hline
HFL              & 1620.74 & 2249.22 & 2983.35 & 3508.61 \\
\hline
\textbf{$\mathsf{HierSFL}$}     & \textbf{1510.35} & \textbf{2145.48} & \textbf{2745.51} & \textbf{3334.84} \\
\hline
\end{tabular}
\label{tab:mnist_training_time}
\end{table}
$\mathsf{HierSFL}$ consistently outperforms HFL and SFL in training speed for both CIFAR-10 and MNIST. The advantage arises from $\mathsf{HierSFL}$ effectively balancing client workloads and minimizing aggregation computations through MESs. As the number of global aggregation rounds increases, the differences in training times among $\mathsf{HierSFL}$, HFL, and SFL become more pronounced. For example, at 25 rounds, $\mathsf{HierSFL}$ finishes training in 3502.75 seconds, while HFL and SFL require 3690.21 and 3868.33 seconds, respectively. This trend holds in MNIST as well, confirming that $\mathsf{HierSFL}$ is a more efficient framework for training on these datasets.

\section{Conclusion}
\label{Section:Conclusion}

The paper has introduced a novel $\mathsf{HierSFL}$ framework that has addressed challenges in MES and FL contexts, assisting clients with limited resources in model training participation. To achieve the goal of enabling clients with limited resources to contribute to model training, $\mathsf{HierSFL}$ has utilized multiple MESs for partial model aggregation while minimizing the training time and energy consumption. Empirical assessments have also revealed the framework's advantage over established two-layer FL architectures in balancing communication and computing. Notably, LDP was incorporated to enhance the confidentiality of local model parameters during the synchronization process. For the future work, adaptive privacy budgets are expected to be integrated into $\mathsf{HierSFL}$, reducing computational costs when adding LDP noise to local client models.

\balance
\bibliography{Reference}
\bibliographystyle{IEEEtran}

\end{document}